\documentclass[%
 reprint,
 amsmath,amssymb,
 aps,
pra,
]{revtex4-1}

\usepackage{graphicx}
\usepackage{dcolumn}
\usepackage{bm}
\usepackage{enumitem}
\usepackage{braket}
\usepackage{tikz}

\usepackage{algorithm}
\usepackage[noend]{algpseudocode}

\begin{document}


\title{Quantum Control with Measurements and Quantum Zeno Dynamics}

\author{J. J. W. H. S\o rensen}
\author{M. Dalgaard}
\author{A. H. Kiilerich}
\author{K. M\o lmer}
\author{J. F. Sherson}
\email{sherson@phys.au.dk}
\affiliation{Department of Physics and Astronomy, Aarhus University, Ny Munkegade 120, 8000 Aarhus C, Denmark}
\date{\today}

\begin{abstract}
We introduce an efficient iterative method to prepare a target state in Hilbert spaces with high dimensionality using a combination of unitary evolution, measurements, and quantum Zeno dynamics. The latter confines the evolution within Zeno subspaces of decreasing size. This gives an exponential speed up relative to the case of states evolving in the full Hilbert space between projective measurements. We demonstrate our approach on the control problem of rapidly transferring a superfluid into the Mott insulator in the Bose-Hubbard model. We discuss the general applicability of the method by preparing arbitrary superpositions with random Hamiltonians.
\\
\end{abstract}

\maketitle

\section{Introduction}
Preparation of specific target quantum states is a prerequisite for, e.g., control of qubits, quantum computation, quantum metrology, and simulation of novel matter phases \cite{brif2010control,norris2012enhanced,lewenstein2007ultracold}. This type of control is typically achieved by manipulating the unitary dynamics using quantum optimal control theory \cite{glaser2015training,werschnik2007quantum,sorensen2018approaching,sorensen2018gradient}.

Although quantum optimal control theory has been applied successfully in several systems, it is still challenging to control many-body systems such as ultracold atoms. A paradigmatic and experimentally relevant example is to transfer a state from the superfluid phase into the Mott insulator phase \cite{braun2015emergence, BH_experimental_realization,doria2011optimal}, which is the starting point for applications such as performing quantum logic gate operations \cite{two_qubit_quantum_gate_by_cold_controlled_collisons, quantum_gate_BH_via_collisions,laser_induced_quantum_gate_operations,FastQuantumGates,Toffili_gate_1D_lattice}, quantum simulations \cite{quantum_simulator}, and single atom transistors \cite{single_atom_transistor}. This transfer is difficult since the adiabatic time scales diverge close to the phase transition where the gap to the excited state closes in an infinite system \cite{cucchietti2007dynamics}. There have been attempts to numerically optimize the transfer using optimal control theory and adiabatic ramp shapes \cite{ramping_optimization_zakrzewski,doria2011optimal,van2016optimal}, and the transition has been studied and optimized experimentally \cite{rosi2013fast,braun2015emergence,van2016optimal}.

An alternative to unitary control is to steer the dynamics using the backaction associated with quantum measurements \cite{jacobs2010feedback,blok2014manipulating,roch2014observation,pechen2006quantum,ashhab2010control,tanaka2012robust}. For instance, by measuring a sequence of observables in spin systems, it is possible to prepare desirable local properties as well as long range correlations \cite{hauke2013quantum}. Measurement-based control of many-body systems requires inclusion of the quantum back-action in the modeling. Initial steps in this direction have been taken in Refs.~\cite{mazzucchi2016quantum,mazzucchi2016collective,kozlowski2017quantum}, where it is shown that collective weak measurements of the on-site densities or coherences in an optical lattice can be used to engineer correlated tunneling and long-range entanglement. The weak measurements confine the system to distinct \textit{Zeno-subspaces} defined by an effective non-Hermitian Hamiltonian, governing the evolution of the monitored system \cite{kozlowski2017quantum}, and Raman-like transitions may be observed between these subspaces \cite{elliott2016quantum,kozlowski2016non}.
\begin{figure*}[t]
\centering
\includegraphics[width=\textwidth]{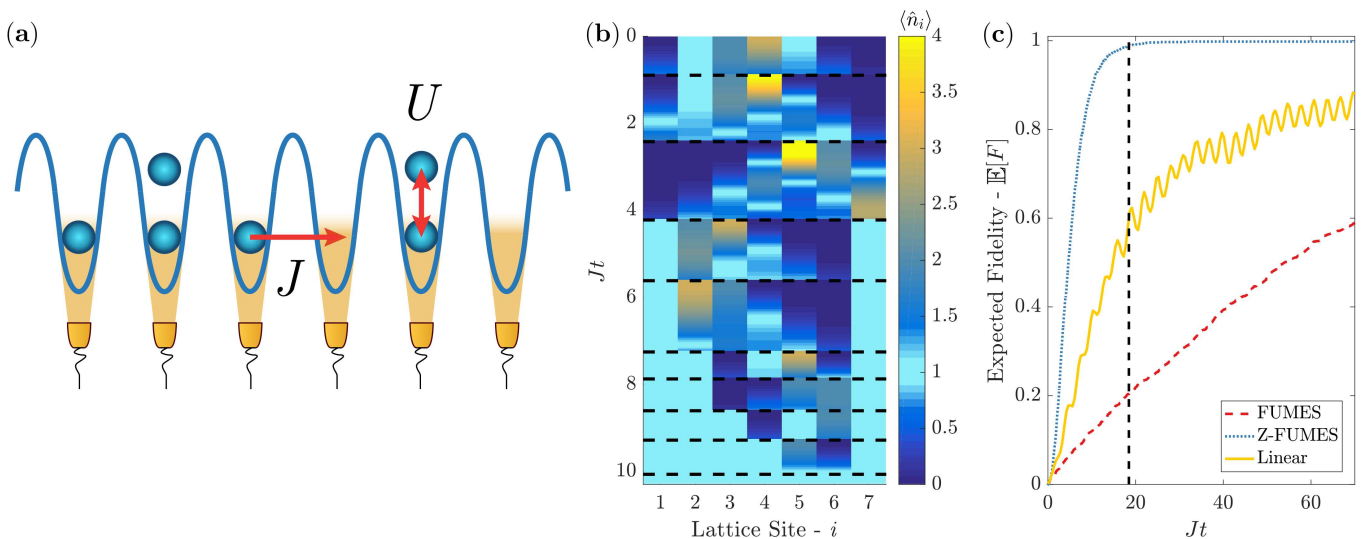} %
\caption{(Color Online) (\textbf{a}) Bose-Hubbard model~(\ref{BHModel}) and the measurements of the single-site operators ($ \hat{n}_i $). (\textbf{b}) A single Z-FUMES trajectory starting from the superfluid state and reaching the Mott state for a lattice with $N=7$ particles and $L=7$ sites. A measurement is performed at each of the dashed, horizontal lines. The color map tracks the populations, $\langle \hat{n}_i\rangle$ on the individual sites. (\textbf{c}) Fidelity for FUMES and Z-FUMES averaged over 1000 simulated trajectories for $N=L=7$. The solid (yellow) curve shows the deterministic fidelity obtained using a linear ramp without measurements from $U/J=0$ at $Jt$ zero to $U/J=30$ at time $Jt$. The rapid, coherent oscillations are due to the non-adiabatic excitation of the system. The vertical line indicates that Z-FUMES reaches $F=0.99$ at $Jt=18.5$.}
\label{fig:BHCompare}
\end{figure*}

Here we discuss combinations of unitary and measurement-based dynamics that may offer better control strategies. Our focus will be on preparation of a target state. For this purpose, one previously proposed control strategy is fixed unitary evolution and measurements (FUMES) \cite{pedersen2014many}. In FUMES it is the timing of the measurements, rather than the unitary dynamics, which is optimized. This means that the unitary dynamics is given by a fixed static Hamiltonian while measurements, attempting to project the system into the target state, are performed at the times with highest success probability. In Ref. \cite{pedersen2014many} it was shown that FUMES is competitive with other types of measurement-based control schemes such as multiple evenly distributed observables (MEDO) \cite{roa2007quantum} and mutually unbiased measurements (MUM) \cite{roa2006measurement}.

A projective measurement on a many-body system is typically realized by many individual (local) measurements. Even if the full projective measurement fails to produce the desired outcome, some of the individual measurements might still have succeeded. Despite exhibiting superior performance to MEDO and MUM, the FUMES strategy suffers from the drawback that it cannot maintain these partial successes. 
In this paper we propose to employ quantum Zeno dynamics to improve the FUMES strategy by freezing the state components prepared by each partial success \cite{misra1977zeno}. This effectively confines the unitary dynamics to smaller Zeno subspaces similar to what was found in Refs.~\cite{elliott2016quantum,kozlowski2016non}. We demonstrate that this gives an exponential speed-up relative to FUMES.   

In Sec.~\ref{sec:BHModel} we introduce FUMES and Z-FUMES as methods for preparing a Mott insulator starting from the superfluid. In Sec.~\ref{sec:weakMeasure}, we analyze the performance of Z-FUMES in a more realistic setting using continuous homodyne measurements rather than projective measurements on the Bose-Hubbard model. In Sec.~\ref{sec:RandHam}, we demonstrate the general applicability of Z-FUMES by simulating the preparation of arbitrary states using random Hamiltonians and measurements. Section \ref{sec:con} concludes the paper.

\section{Control of the Bose-Hubbard Model} \label{sec:BHModel}
\subsection{Control problem}
Ultracold bosons in a one-dimensional optical lattice are described by the Bose-Hubbard model
\begin{equation}
\hat{H}=-J\sum_{j=1}^{L-1}\Bigl(\hat{a}_j^\dag \hat{a}_{j+1}+\hat{a}_{j+1}^\dag \hat{a}_j\Bigr) + \frac{U}{2}\sum_{j=1}^L \hat{n}_j(\hat{n}_j-1), \label{BHModel}
\end{equation}
where \textit{J} is the tunneling rate, \textit{U} is the on-site interaction and \textit{L} is the number of sites, see Fig. 1\textbf{a}. $\hat{a}_j$ and $\hat{n}_j$ are the annihilation and number operators for particles on the \textit{j}'th site respectively. Here we assume unit filling, i.e., that the number of particles matches the number of sites ($N=L$). For $U=0$ or $J=0$ the ground state of the system has different quantum phases, which are the superfluid and Mott insulator respectively \cite{bloch2008many}. In this paper we refer to the ground state of the $U=0$ Hamiltonian as the superfluid state. As discussed below, our method may be applied to prepare general many-body correlations. However, we first focus on the intuitive but experimentally relevant case of transferring a system from an initial superfluid state into the Mott state $|\text{Mott}\rangle = |1,1,...,1\rangle$. The quality of this transfer is quantified by the fidelity $F=|\langle \text{Mott}|\psi\rangle|^2$ where $\psi$ is the state at the end of the control protocol.

Recent advances in single atom detection have made it feasible to image single atoms in both optical lattices \cite{sherson2010single,greif2016site,parsons2015site} and free space \cite{bergschneider2018spin}. Including Raman sideband cooling allows the atoms to be detected without additional heating \cite{greif2016site,patil2015measurement,patil2014nondestructive}. Dispersive imagining of single atoms in an optical lattice has also recently been realized using the Faraday effect \cite{yamamoto2017site}.

In principle this enables the implementation of quantum non-demolition measurements of the local atom-numbers, thus providing access to the set of observables $\{\hat{n}_1,\hat{n}_2,...,\hat{n}_L\}$ for both the projective measurement and the quantum Zeno dynamics, see Fig.~\ref{fig:BHCompare}(\textbf{a}). A simultaneous measurement of all observables collapses the state into a Fock-state $|n_1,n_2,...,n_L \rangle$ with $\sum_j n_j=N$. This type of system can be controlled using the FUMES control strategy introduced in Ref.~\cite{pedersen2014many}. In FUMES the system is projected into a Fock-state by simultaneous measurements of all the $\hat{n}_i$ operators. This type of measurement will either succeed by projecting into the Mott state or fail by a projection into another Fock-state. For a non-zero \textit{J} the Fock-states are not eigenstates of the model~(\ref{BHModel}), which implies that if the measurement fails then the subsequent unitary dynamics drives the system out of the projected Fock-state. Hence, at later times there is again a non-zero probability of projecting into the target state. If the projections occur at arbitary times then there is only a low probability of success \cite{burgarth2007quantum}. In FUMES this probability is improved by only measuring at peaks in the fidelity above some preset threshold. In an experimental setting, these peaks can be calculated prior to the experiment by solving the deterministic Schr\" odinger equation.
\begin{figure*}[t]
\begin{minipage}[t]{.48\textwidth}
	\includegraphics[width=0.93\textwidth]{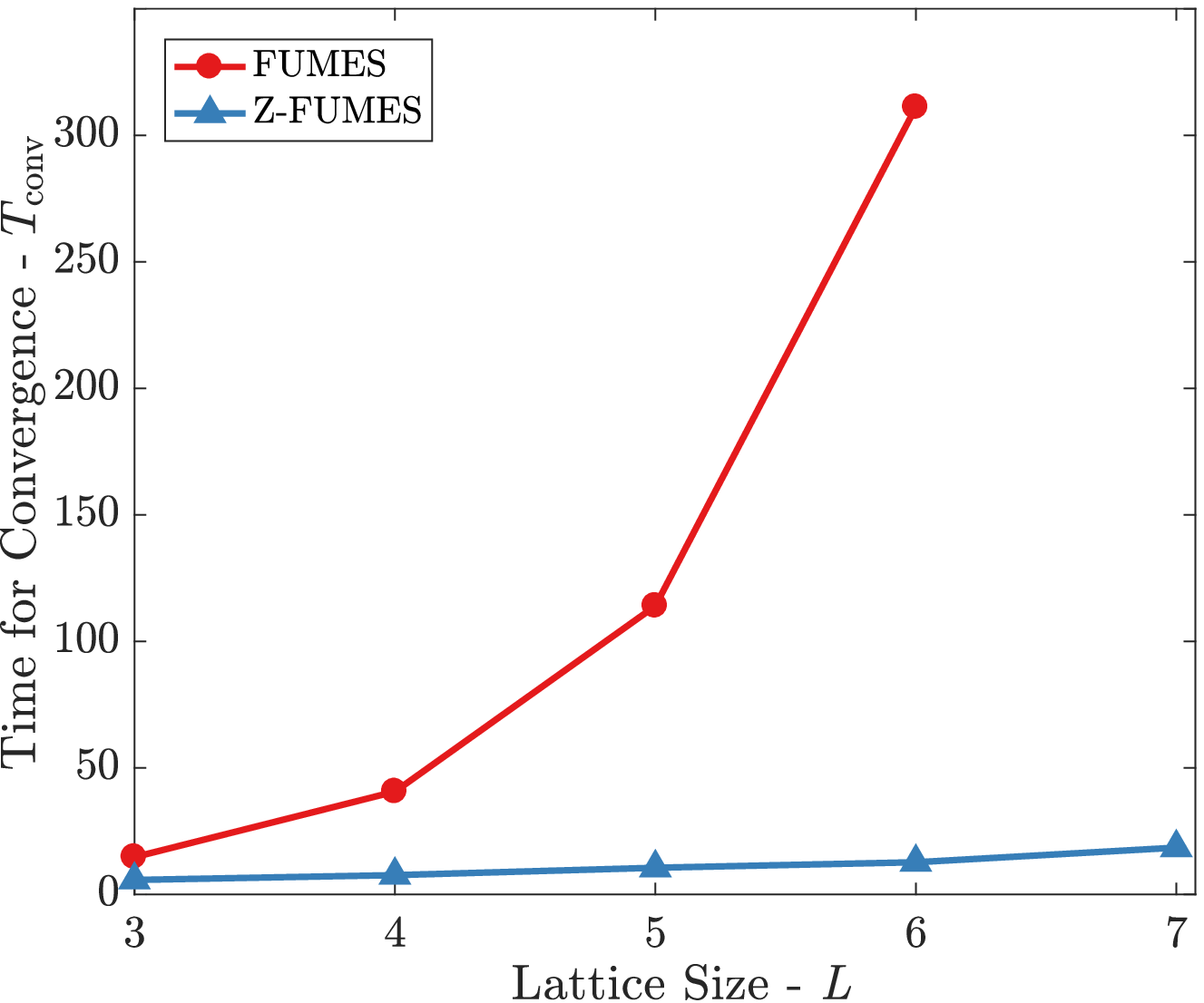}
	\caption{(Color online) Expected $Jt$ needed to reach a fidelity above $0.99$ as a function of the lattice size for FUMES and Z-FUMES simulated under the Bose Hubbard model~(\ref{BHModel}). The results are averaged over 1000 trajectories for each value of \textit{L}.}
	\label{fig:BHScaling}
\end{minipage}
\;
\begin{minipage}[t]{.48\textwidth}
  \includegraphics[width=0.93\textwidth]{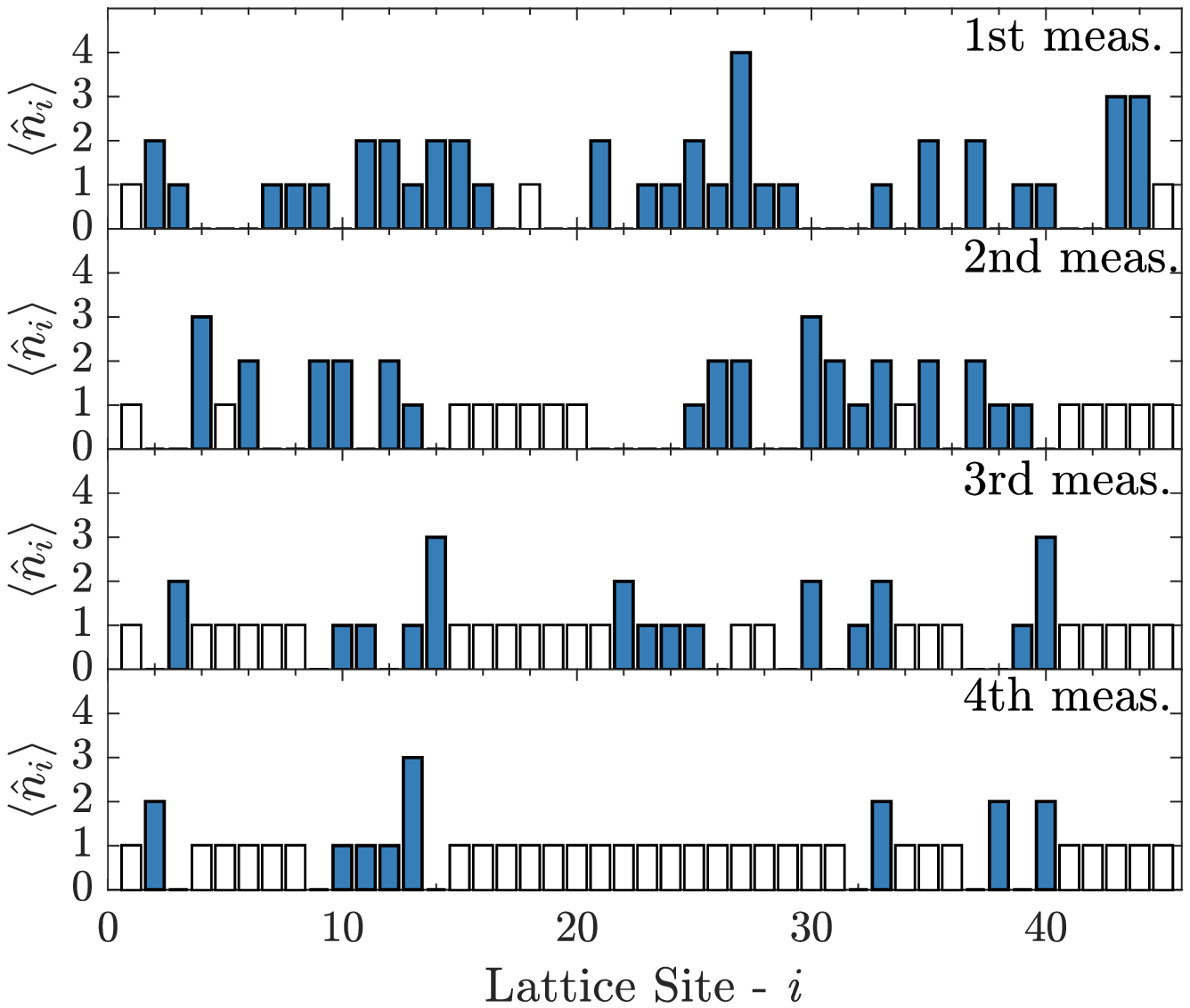}
  \caption{First four measurements of a toy model simulation of Z-FUMES assuming a complete reshuffling between measurements. Zeno locked sites are shown with white. The distributions of particles, $(n_1,n_2,\dots,n_L)$ after a measurement is sampled from Eq.~(\ref{PDF}) for each sublattice with $L=L_{\text{sub}}$, where $L_{\text{sub}}$ is the sublattice length.}
  \label{fig:repeatedMeasurements}
\end{minipage}
\end{figure*}
 FUMES is problematic in the sense that for a large lattice it becomes exponentially improbable to project directly into the Mott state. In order to remedy this effect, we propose Z-FUMES, which is a combination of FUMES and quantum Zeno dynamics \cite{misra1977zeno}. Quantum Zeno dynamics in optical lattices have been reported experimentally in Refs. \cite{patil2015measurement,patil2014nondestructive}. Although a projective measurement of all the sites may not have reached the Mott state, some of the individual sites may still have the desired unit occupancy. In order to prevent these particles from tunneling away we propose to trap them using quantum Zeno dynamics, i.e., by performing rapid repeated measurements of the on-site number operator \cite{misra1977zeno}. \textit{Zeno-locking} the number of particles on a site does not only ensure the correct occupancy, it also prevents particles from tunneling across that site. This implies that locked sites effectively decouple the lattice into smaller parts. However, the Mott state can only be reached if each of these sublattices contain the correct number of particles. Hence, we should only Zeno-lock a given site whenever it contains a single particle \textit{and} the right and left sublattices have matching numbers of sites and particles.

We compare FUMES and Z-FUMES for creating a Mott state in this model for a system with size $N=L=7$. Between the measurements the evolution of the state is governed by the Hamiltonian~(\ref{BHModel}) with $U/J=0$. The on-site density ($\langle \hat{n}_i \rangle$) during a Z-FUMES trajectory is shown in Fig. \ref{fig:BHCompare}(\textbf{b}). The state is initially in the superfluid state and discrete changes in the density are introduced by projective measurements of the on-site density at specific times marked by dashed lines. At the time $Jt=4$, the two outer sites have been Zeno locked, creating a sublattice of length five. The edges of this sublattice are gradually Zeno locked in this trajectory, and after about $Jt=10$, the system has converged to the Mott insulator state with $\langle \hat{n}_i \rangle=1$ for $i=1,2,...,7$.  The gradual locking in the proper subspaces is the reason Z-FUMES converges faster than FUMES.

In Fig.~\ref{fig:BHCompare}(\textbf{c}) the mean fidelity as a function of $Jt$ is shown for both FUMES and Z-FUMES. The curves are obtained by averaging the results of 1000 simulated trajectories. The figure shows that Z-FUMES reaches an expected unit fidelity after about $Jt=20$ while after $Jt=70$ FUMES still only has a success rate of $60\%$. 
For comparison we also show the fidelity after a unitary linear ramp of the interatomic interaction strength from $U/J=0$ to $U/J=30$ during the same time interval but in the absence of any measurements. For each value of $Jt$, the ramp is thus performed with a different speed. The solid (yellow) line in Fig.~\ref{fig:BHCompare}(\textbf{c}) shows the final fidelity as a function of the total ramp time. For higher values of $Jt$ the transfer is adiabatic and the fidelity will approach unity. The curve further exhibits characteristic rapid, coherent oscillations, which are due to the energy differences between populated eigenstates during the transfer.
Although the fidelity from the linear ramp lies higher than FUMES on the curve, one should remember that the linear ramp never reaches a pure Mott state whereas FUMES leads to formation of the pure Mott state in $60\%$ of the simulated runs. Z-FUMES clearly performs better than both the linear ramp and FUMES.

\subsection{Scaling with system size}
In this part, we discuss how FUMES and Z-FUMES scale with the lattice size in the Bose-Hubbard model~(\ref{BHModel}). For this purpose, we define $T_{\text{conv}}$ as the time where the mean fidelity reaches $\mathbb{E}[F]=0.99$. The scaling of this quantity with the lattice size $L$ is illustrated in Fig.~\ref{fig:BHScaling}
where $T_{\text{conv}}$ is averaged over 1000 simulations for each value of \textit{L}. FUMES scales poorly with the lattice size as it becomes exponentially improbable to project the system into the Mott state. The improved scaling in Z-FUMES is due to the fact that each time a site is locked, the lattice is divided into smaller sublattices each with a higher probability of measuring the desired outcome in subsequent measurements.

Due to the exponential growth of the Hilbert space, it is not possible to simulate Eq.~(\ref{BHModel}) for large systems. However, it is possible to perform a toy model analysis by assuming a complete reshuffling within each sublattice after a measurement. The probability for a particle distribution $\mathbf{n}=(n_1,n_2,...,n_L)$ is given by the multinomial expression,
\begin{equation}
P(\mathbf{n})=\frac{1}{n_1!n_2! \cdots n_L!} \frac{L!}{L^L}. \label{PDF}
\end{equation}
For FUMES the mean number of measurements needed for convergence $M^F$ may be estimated directly from Eq.~(\ref{PDF}) (see Appendix)
\begin{equation}
M^F \simeq \frac{e^L}{\sqrt{2 \pi L}}. \label{eq:MF}
\end{equation}

In each iteration of the toy model simulation of Z-FUMES the outcome of a measurement on each sublattice $L_{\text{sub}}$ is sampled from Eq.~(\ref{PDF}) with $L=L_{\text{sub}}$.  The first few iterations of one realization within this toy model simulation are shown in Fig.~\ref{fig:repeatedMeasurements} where the bars represent the number of atoms on a given site and the white bars show the Zeno locked sites. Here, after just four measurements most of the sites have been Zeno locked.

A conservative estimate of the number of measurements needed for convergence for Z-FUMES $M^Z$ is explained in the Appendix by assuming a uniform probability distribution among the Fock states,
\begin{equation}
M^Z \lesssim 16 \sqrt{\frac{L}{\pi}}. \label{eq:MZ}
\end{equation}
The non-exponential scaling clearly demonstrates the power of continually dividing the system into ever smaller decoupled sublattices. Equations ~\eqref{eq:MZ} and \eqref{eq:MF} clearly show that Z-FUMES achieves an exponential speed up compared to FUMES, resulting in the favorable scaling seen in Fig.~\ref{fig:BHScaling}. The trajectory in Fig.~\ref{fig:BHCompare}(\textbf{b}) simulated with Eq.~(\ref{BHModel}) used nine measurements for convergence whereas Eq.~\eqref{eq:MZ} estimates $M^Z\leq 24$ showing that the estimate is not tight.

\section{Continuous Measurements} \label{sec:weakMeasure}
\begin{figure}
\includegraphics[width=\columnwidth]{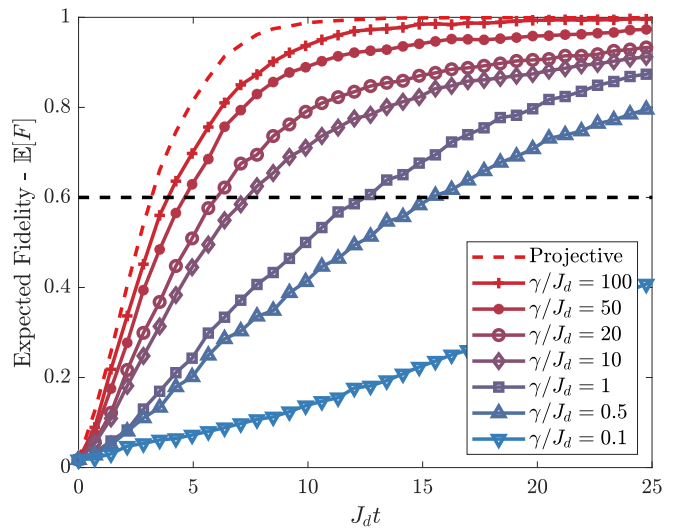}
\caption{(Color Online) Expected fidelity in Z-FUMES averaged over 1000 trajectories simulated using continuous homodyne measurements. The simulations are made for different measurement strengths $\gamma$ and a lattice with $L=5$ sites. The dashed curve is not simulated using continuous homodyne measurements but with unitary dynamics and projective measurements as in Fig.~\ref{fig:BHCompare}(\textbf{c}). The horizontal dashed lines indicates 60\% expected fidelity.}
\label{fig:WeakFUMES}
\end{figure}
In the previous sections we have assumed that the measurements occur instantaneously. This is a valid assumption when the typical duration of a measurement is short compared to the dynamics of the measured system \cite{jacobs2006straightforward}. However, this is not generally true in practical applications with finite interaction strengths where the measurement record $I(t)$ is continuous \cite{yamamoto2017site}. 
As a step towards experimental realizability we investigate the performance of Z-FUMES with weak continuous measurements. 
\begin{figure*}[t]
\begin{minipage}[t]{.48\textwidth}
	\centering
\begin{tikzpicture}[scale = 0.95]
\draw [fill=red!20, opacity = 0.5, ultra thick, rotate = 0] (0,-1.3cm) circle [radius = 2.05cm];
\draw [fill=blue!20, opacity = 0.5, ultra thick, rotate = 120] (0,-1.3cm) circle [radius = 2.05cm];
\draw [fill=green!20, opacity = 0.5, ultra thick, rotate = 240] (0,-1.3cm) circle [radius = 2.05cm];
\node at (0,0) {$\ket{1,1,1}$}; 
\node at (0,1.5) {$\ket{1,1,3}$};
\node at (0,1.0) {$\ket{1,1,2}$};
\node at (1.25,-0.5) {$\ket{2,1,1}$};
\node at (1.4,-1.0) {$\ket{3,1,1}$};
\node at (-1.2,-0.5) {$\ket{1,2,1}$};
\node at (-1.35,-1.0) {$\ket{1,3,1}$};
\node at (-1.2,2.2) {$\ket{1,3,3}$};
\node at (-2.1,1.5) {$\ket{1,3,2}$};
\node at (-2.5,0.8) {$\ket{1,2,3}$};
\node at (-2.3,0.1) {$\ket{1,2,2}$};
\node at (1.2,2.2) {$\ket{3,1,3}$};
\node at (2.1,1.5) {$\ket{3,1,2}$};
\node at (2.5,0.8) {$\ket{2,1,3}$};
\node at (2.3,0.1) {$\ket{2,1,2}$};
\node at (-1.2,-1.85) {$\ket{3,3,1}$};
\node at (-0.65,-2.65) {$\ket{3,2,1}$};
\node at (0.65,-2.65) {$\ket{2,3,1}$};
\node at (1.2,-1.85) {$\ket{2,2,1}$};
\node at (0.0,3.0) {$\braket{\hat{Q}_2} = 1 $};
\draw [-] (0.27, 2.87) -- (0.4, 2.6);
\node at (2.2,-3.2) {$\braket{\hat{Q}_3} = 1 $};
\draw [-] (2.4, -3.1) -- (1.8, -2.35);
\node at (-2.9,-1.7) {$\braket{\hat{Q}_1} = 1$};
\draw [-] (-2.6, -1.6) -- (-2.2, -1.1);
\end{tikzpicture}
	\caption{(Color Online) Illustration of how the $q_j$ are distributed in a general context. The circles show the Zeno subspace corresponding to a value of $q_j^{(i)}$ for a particular $\hat{Q}_i$. Each Zeno-subspace contains a number of the basis states $|q_1,q_2,...,q_L\rangle$. A simultaneous measurement of all the $\hat{Q}_i$ observables yields a single state projection, which in this illustration is into the target state $\ket{\psi_{\mathrm{target}}}=|1,1,1\rangle$. In this example we have $B=L=3$.}
	\label{fig:qjConstruction}
\end{minipage}
\quad
\begin{minipage}[t]{.48\textwidth}
  \includegraphics[width=\textwidth]{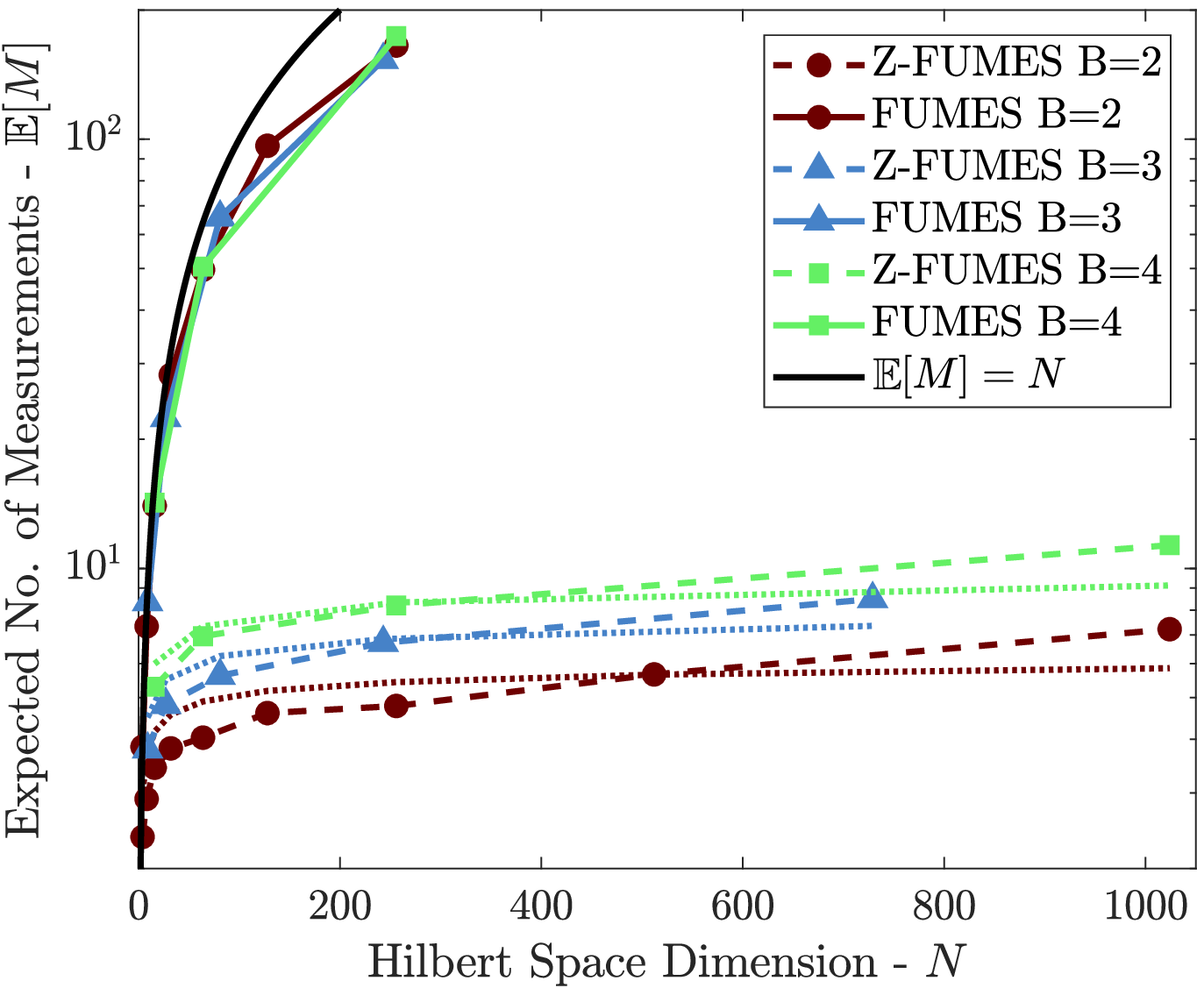}
  \caption{(Color Online) Expected number of measurements needed for convergence $\mathbb{E}[M]$ for FUMES and Z-FUMES for systems with different number of measurement operators \textit{L} and outcomes \textit{B} - see legend. The dotted lines show the estimate from Eq.~\eqref{generalMZ}. Each point is averaged over 1000 random Hamiltonians and initial states. The Hilbert space dimension is $N=B^L$ meaning that a larger \textit{N} corresponds to more measurement operators.}
  \label{fig:FUMESAndZeno}
\end{minipage}
\end{figure*}

Due to the backaction of the continuous measurements, the dynamics obeys a stochastic Schr\" odinger equation, which can be understood as the time evolution conditioned on the stream of measurement results in $I(t)$ \cite{jacobs2006straightforward,wiseman2009quantum}. 
At each instant in time, $I(t)$ is dominated by stochastic noise, resulting in a state $|\psi(t)\rangle$, which evolves in a random manner. This time evolution exhibits quantum jumps if measurement outcomes occur at discrete points in time as in photo detection while for example homodyne detection leads to a diffusive trajectory \cite{wiseman2009quantum}. Quantum jump trajectories have previously been studied in the context of the Bose Hubbard model \cite{mazzucchi2016quantum,kozlowski2016non,mazzucchi2016collective,kozlowski2017quantum,mekhov2009quantum,mekhov2007probing}.
Here we consider diffusion-type measurements which imply a stochastic Schr\" odinger equation of the form
	\begin{equation}
	\text{d} \ket{\psi(t)}
	= \text{d}t
	\left[
	-i \hat{H}(t)
	+
	\sum_j
	-\frac{\gamma_j}{2} \hat{c}_j^{\dagger} \hat{c}_j
	+ I_j(t) \hat{c}_j
	\right]\ket{\psi(t)}, \label{SSE}
	\end{equation}
where the $\hat{c}_j = \hat{n}_j$ are the measurement operators and the $\gamma_j$ corresponding measurement strengths, which determine the rate at which information is extracted \cite{Carmichael1993,wiseman2009quantum}. In the examples studied here, we assume uniform strengths $\gamma_j \equiv \gamma$. A measurement becomes projective in the limit $\gamma \rightarrow \infty$ \cite{jacobs2006straightforward}. Note that Eq.~(\ref{SSE}) does not preserve the normalization, which is instead imposed explicitly in each time step. The measurement record $I_j(t)$ for the \textit{j}'th detector reflects the current state of the system,
	\begin{align}
	I_j(t) = \gamma_j \braket{\hat{c}_j^{\dagger} + \hat{c}_j} (t) + \sqrt{\gamma_j} \,	\xi_j(t),
	\end{align}
where the $\xi_j(t) = \text{d}W_j(t)/\text{d}t $ are infinitesimal Wiener increments, representing white noise in the detection setup. Integration of the record allows one to determine the outcome of a measurement for large values of $\gamma$. In order to investigate the performance of FUMES and Z-FUMES at low values of $\gamma$, we assume that it is possible to quench the lattice such that $J(t)=J_d$ with $U(t)=0$ in the absence of measurements and $J(t)=0$ with $U(t)=J_d$ during measurements, where $J_d$ is the characteristic dynamical energy scale. After a measurement some of the particles may be trapped by the Zeno effect using sufficiently strong measurements to suppress tunneling. In our simulations we used $\gamma_j/J_d=1000$ for Zeno locking of the particle on site \textit{j}.

In Fig.~\ref{fig:WeakFUMES}, Z-FUMES is investigated for different values of the measurement strength in a system with $L=5$ lattice sites. The figure also shows the expected fidelity for Z-FUMES simulated using discrete projective measurements as in Fig.~\ref{fig:BHCompare}(\textbf{c}). As expected, the fidelity of the continuous measurement scheme converges towards that corresponding to projective measurements as the measurement strength becomes large. At the time $J_dt=15$ more than 60\% of the trajectories have converged with $\gamma/J_d = 0.5$, which means that Z-FUMES is effective even with moderate values of the measurement strength.

\section{General application of Z-FUMES} \label{sec:RandHam}
In this section we show that FUMES and Z-FUMES can also be applied to prepare an arbitrary target state of a system, almost independently of its Hamiltonian evolution. In the last section we gave the example of preparing an individual Fock state by measuring the on-site populations. If the measurements were instead performed in the Fourier basis (corresponding to momentum eigenstates) then many-body states with exotic phase correlations could be prepared \cite{pedersen2014many,kozlowski2017quantum,hauke2013quantum}. To demonstrate this general applicability in an unbiased way, we here assume random Hamiltonians drawn from the Gaussian unitary ensemble which ensures a uniform distribution in the space of Hamiltonians with a particular dynamical time-scale \cite{zirnbauer1996riemannian}. This type of random Hamiltonians is used to model, e.g., chaotic systems with one and many particles \cite{stockmann2000quantum}.

We assume the system can be manipulated through the backaction from measuring a set of commuting observables $\{\hat{Q}_1,\hat{Q}_2,...,\hat{Q}_L\}$, $[\hat{Q}_i,\hat{Q}_j]=0$. The joint eigenstates of the measurement operators define a set of orthonormal states $|q_1,q_2,...,q_L\rangle$ and the target state $|\psi_{\text{target}}\rangle$ must be one of these eigenstates. A measurement of all $\hat{Q}_i$ observables must uniquely determine a $|q_1,q_2,...,q_L\rangle$ state. Each measurement operator may be written as a linear combination $\hat{Q}_i = \sum_{i} q_j^{(i)} |...,q_j^{(i)},...\rangle \langle ...,q_j^{(i)},...|$. The eigenvalues $q_j^{(i)}$ should be constructed with a large degeneracy such that they define different subspaces; see Fig. \ref{fig:qjConstruction}. A measurement of a $\hat{Q}_i$ is successful if it measures the same $q_j^{(i)}$ as for the target state. Zeno-locking this value confines the time evolution to a smaller subspace containing the target state. The subsequent time evolution within this subspace is
\begin{equation}
\hat{U}_Z(\Delta t) = \exp\Bigl(-i\hat{P}\hat{H}\hat{P} \Delta t\Bigr), \label{ZenoFormula}
\end{equation}
where $\hat{P}$ is the projector on the locked Zeno-subspace \cite{facchi2008quantum}. A subspace should only be Zeno-locked if there is a sufficient coupling between the current state and the target state within that subspace, i.e. $|\langle \psi_{\text{target}}|\hat{U}_Z(\Delta t)|\psi(t)\rangle|>\epsilon_o$ where $\epsilon_o$ is a predefined threshold.

These ideas constitute a direct generalization of the Bose-Hubbard control scheme discussed in the previous sections. There the orthonormal basis consists of the Fock-states and the individual on-site number operators can be written as linear combinations of projectors on these states. We have for instance, that the on-site density operator for the \textit{i}'th site is $\hat{n}_i = \sum_i n_j^{(i)} |...,n_j^{(i)},...\rangle\langle ...,n_j^{(i)},...|$. The eigenspectrum of the number operator on a single site is clearly degenerate. Measuring the $\hat{n}_i$ operators one-by-one gradually leads to a collapse onto a single Fock-state. The condition of only Zeno-locking in subspaces with sufficient coupling between the current state and the target state corresponds to exclusively locking sites with matching numbers of particles and sites on the left and right sublattices.

We have applied Z-FUMES to perform general state transfers using random Hamiltonians. In order to show the average behavior, we have performed calculations for 1000 different random Hamiltonians for each size of the Hilbert space. We assume access to \textit{L} different measurement operators each with \textit{B} different outcomes giving a Hilbert space of size $N=B^L$. Fig.~\ref{fig:FUMESAndZeno} compares the number of measurement needed for convergence $\mathbb{E}[M]$ for different number of measurement operators (\textit{L}) and number of outcomes (\textit{B}). As in the case of the Bose-Hubbard model, the Z-FUMES curves converge much faster than FUMES, requiring about two orders of magnitude fewer measurements. All Z-FUMES curves show a similar rate of convergence despite the system size differing by an order of magnitude. 

The scaling with the system size may be understood by a simple model. Assuming that all measurement outcomes are equally likely, the probability of a measured $\hat{Q}_i$ producing the desired outcome is $B^{-1}$. If the measurements are independent, the mean number of locked observables is $K/B$ when measuring \textit{K} observables. The number of measurements needed to converge may then be estimated by summing the average time for locking each operator
\begin{equation}
M^Z = B \sum_{K=1}^L \frac{1}{K} \simeq B \ln L. \label{generalMZ}
\end{equation}
This logarithmic scaling is slower than the square root found for the Bose-Hubbard model in Eq.~(\ref{eq:MZ}). In the Bose-Hubbard model there are fewer states in the Hilbert space, but only a few states are lockable due to the constraint that the left and right sublattice must have matching number of particles and sites. In combination this gives a lower probability for locking a site than in the unconstrained case. The value in Eq.~(\ref{generalMZ}) is compared with the simulations using random Hamiltonians in Fig.~\ref{fig:FUMESAndZeno}. Fig.~\ref{fig:FUMESAndZeno} shows that it is favorable to have fewer possible measurement outcomes, which is also captured by Eq.~(\ref{generalMZ}). For larger systems, Eq.~(\ref{generalMZ}) seems to underestimate the number of measurements needed, which may be due to violation of the assumption that the observables are independent.

In FUMES the system either projects into the target state or not and assuming a uniform distribution the expected number of measurements is $M^F = B^L = N$. This linear scaling is plotted with a black line in Fig.~\ref{fig:FUMESAndZeno}. The FUMES curves follow a linear scaling depending only on \textit{N}, but the slope is lower than unity. We attribute this to the fact that FUMES only performs the measurements at peaks in the fidelity, but also note that the scaling may change for larger Hilbert space dimensions.

\section{Conclusion} \label{sec:con}
We have introduced a protocol, denoted Z-FUMES, for multi-particle state preparation using unitary dynamics and measurement-based control. Our protocol steers the evolution towards a target state by measuring a set of observables. Each observable is Zeno-locked when an appropriate outcome is obtained, which confines the time evolution to gradually shrinking Zeno-subspaces. Z-FUMES gives an exponential speed up compared to other measurement-based control protocols.

We analyzed in detail the preparation of a Mott state using Z-FUMES. Here it is necessary to measure the density on each lattice site, which can be realized experimentally using strong fluorescence imaging combined with Raman side-band cooling as shown in Refs.~\cite{patil2014nondestructive,patil2015measurement}. We demonstrate furthermore that Z-FUMES can be applied for preparing a Mott state in more realistic settings relying on weak continuous rather than projective quantum measurements. We also discussed how to implement Z-FUMES in a more general setting with random Hamiltonians.

In this work we focused on local measurements in order to reach the Mott insulator state which is characterized by local properties. It is also possible to generate strongly correlated states using non-local measurements \cite{hauke2013quantum}. This could be used to engineer exotic quantum state such as Schr\" odinger cats \cite{pedersen2014many} or other states with long-range correlations \cite{mazzucchi2016collective}.

\section{Acknowledgements}
This work has been funded by the European Research Council, the Lundbeck Foundation, the John Templeton Foundation, and the Villum Foundation.

The authors would like to thank Jonathan Satchell for graphical assistance with Fig.~\ref{fig:BHCompare}(\textbf{a}). 

\bibliographystyle{apsrev4-1}
\bibliography{references}

\appendix*
\section{Scaling in FUMES and Z-FUMES}
\begin{figure}[t]
\includegraphics[width=\columnwidth]{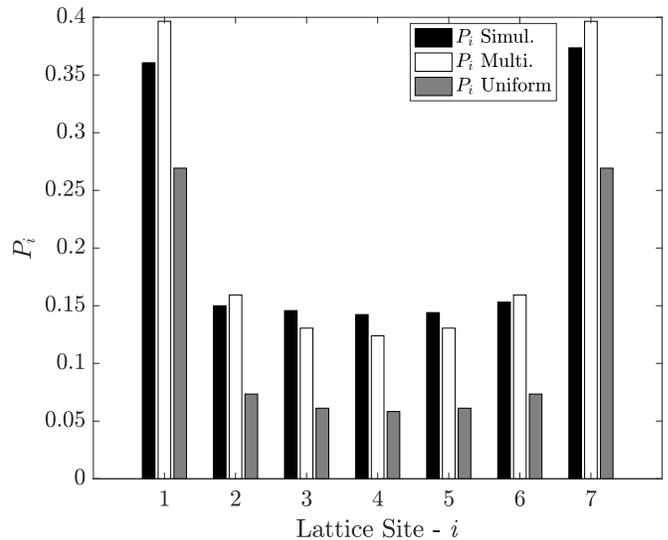}
\caption{The probability distribution $P_i$ for Zeno locking site \textit{i} obtained from the multinomial distribution (white) Eq.~(\ref{PDF}) and uniform distribution (grey) found by summing over all configurations allowing Zeno locking. The black bars show estimated values of $P_i$ by checking which sites could have been Zeno locked from measurements on 600 trajectories using Eq.~(\ref{BHModel}).}
\label{fig:distributions}
\end{figure}

In this appendix we discuss the arguments leading up Eqs.~\eqref{eq:MF} and \eqref{eq:MZ}. The probability for fulfilling the condition to lock the \textit{i}'th site $P_i$ can be found directly from the simulations using Eq.~(\ref{BHModel}). In Fig~\ref{fig:distributions} $P_i$ are estimated from measurements in 600 trajectories for $N=L=7$. The numerical results are compared with the same probabilities found using the multinomial distribution in Eq.~\eqref{PDF} and a uniform distribution where all Fock states are equally likely to be measured. The uniform distribution \textit{underestimates} the probability for locking a specific site, which may be attributed to lockable configurations with a low number of particles on each site being more probable as evident from  Eq.~(\ref{PDF}).

We shall now discuss how to derive an upper estimate for the scaling of Z-FUMES with the number of sites using the uniform distribution. The probabilities in Fig.~\ref{fig:distributions} for the uniform case may be directly found by counting the number of possible configurations allowing a site to be Zeno locked
\begin{equation}
P_i = \frac{C(i-1)C(L-i)}{C(L)}. \label{Pi}
\end{equation}
Here we assume unit filling $L=N$ and $C(N)=(2N)!/2 N!^2$ is the number of ways to distribute \textit{N} particles among \textit{N} sites. In the special cases of $i=1$ or $i=L$ Eq.~(\ref{Pi}) becomes $P_i=L/2(2L-1)$. The limit, $1 \ll i \ll L$ of Eq.~(\ref{Pi}) can be investigated using Stirling's approximation
\begin{equation}
P_i \simeq \frac{1}{8} \sqrt{\frac{L}{\pi(i-1)(L-i)}}. \label{iscaling}
\end{equation}
Seeking an upper estimate on the number of measurements needed for convergence, we compute the average probability for locking a site from Eq.~(\ref{iscaling})
\begin{align}
P_{\text{avg}} \equiv \frac{1}{L}\sum_{i=1}^L P_i &\simeq \frac{1}{2L-1} \nonumber \\ 
&+ \frac{1}{8L} \int_2^{L-1} \sqrt{\frac{L}{\pi(i-1)(L-i)}} \text{d}i 
\end{align}
where the first term is the edge contribution and the discrete sum is approximated by an integral. Calculating the integral and performing a large \textit{L} expansion we obtain
\begin{equation}
P_{\text{avg}} = \frac{1}{8}\sqrt{\frac{\pi}{L}} + \frac{1}{2L} - \frac{1}{2\sqrt{\pi}L^{3/2}} + ... \label{Pavg}
\end{equation}
Assuming that the probability distribution for locking is binomial with success probability $P_{\text{avg}}$, the average number of sites locked in a chain of length \textit{K} is $K P_{\text{avg}}$. An estimate on the number of measurements needed for convergence is found by summing the average time for locking each site
\begin{equation}
M^Z \leq \sum_{K=1}^L \frac{8}{\sqrt{\pi K}} \simeq 16 \sqrt{\frac{L}{\pi}},
\end{equation}
where the discrete sum is approximated by an integral and only the dominant first term from Eq.~(\ref{Pavg}) is included.

In FUMES the collective measurement either projects the state into the target state with $\mathbf{n}=(1,1,...,1)$ or not. The average number of measurements needed for convergence may then be estimated directly from Eq.~(\ref{PDF})
\begin{equation}
M^F = \frac{1}{P(1,1,...,1)} = \frac{L^L}{L!} \simeq \frac{e^L}{\sqrt{2 \pi L}}.
\end{equation}

\end{document}